# Demonstration of the DC-Kerr Effect in Silicon-Rich Nitride

ALEX FRIEDMAN,[1,][*] HANI NEJADRIAHI,[1] RAJAT SHARMA,[1] AND YESHAIAHU FAINMAN[1]

[1] *Department of Electrical & Computer Engineering, University of California, San Diego, 9500 Gilman Drive, La Jolla, CA 92093, USA*
*Corresponding author: amfriedm@eng.ucsd.edu*

**We demonstrate the DC-Kerr effect in PECVD Silicon-rich Nitride (SRN) and use it to demonstrate a third order nonlinear susceptibility, $\chi^{(3)}$, as high as $(6 \pm 0.58) \times 10^{-19}$ m$^2$/v$^2$. We employ spectral shift versus applied voltage measurements in a racetrack ring resonator as a tool by which to characterize the nonlinear susceptibilities of these films. In doing so we demonstrate a $\chi^{(3)}$ larger than that of silicon and argue that PECVD SRN can provide a versatile platform for employing optical phase-shifters while maintain a low thermal budget using a deposition technique readily available in CMOS process flows.**

## Introduction

The search for ever more efficient devices to power the next generation of optical interconnects has long been a driving force behind research in nonlinear optics. Recent studies have focused on exploiting the lithium niobate on insulator platform [1] as well as high-k ferroelectric perovskites such as barium titanate (BaTiO$_3$) [2] in order to realize more efficient electro-optic switches, however, the low refractive index, relative to silicon, and high rf permittivity remain a challenge for these platforms. Furthermore, the strong push towards CMOS compatible fabrication has continued to drive interest in CMOS compatible alternatives for realizing electro-optic switches. The natural choice would be to use a material (or its variant) and an effect that is already available as part of the current silicon photonics platform, such as the plasma dispersion or the DC-Kerr effect. Plasma dispersion-based switching is a commonly utilized technique in the Silicon-On-Insulator (SOI) platform; however, for many applications an ultra-low energy per bit metric is required making all-phase modulation desirable.

An alternative to directly utilizing silicon, to avoid plasma dispersion, is to engineer optical nonlinearities in existing CMOS compatible materials. An attractive candidate for this is non-stoichiometric silicon nitride, and in particular Silicon-rich Nitride. A variety of deposition techniques, including but not limited to Inductively Coupled Plasma Chemical Vapor Deposition (ICP-CVD), Plasma Enhanced Chemical Vapor Deposition (PECVD), and Low-Pressure Chemical Vapor Deposition (LPCVD) have all been used to demonstrate low loss SRN film with enhanced second and third order nonlinear susceptibilities ($\chi^{(2)}$ and $\chi^{(3)}$) [3, 4, 5]. Using these deposition techniques, silicon-rich nitride (SRN) films have demonstrated efficient four-wave mixing where in the case of ultra-silicon-rich nitride $\chi^{(3)}$ coefficients as high as $1.02 \times 10^{-18}$ m$^2$/v$^2$ have been demonstrated [4, 6]. By comparing this to recent work [7] in silicon where DC-Kerr based modulation was demonstrated using a $\chi^{(3)}$ of $2.45 \times 10^{-19}$ m$^2$/v$^2$, the conclusion is that ultra-silicon rich nitride has a variety of highly desirable characteristics for electro-optic switching and the exploration of the DC-kerr effect in this platform deserves consideration.

In this manuscript we will discuss how SRN possess several advantages which make it a strong candidate for practical switching applications. Specifically, we report a PECVD grown SRN film with a refractive index of 3.01896 at 1500nm and experimentally demonstrate a $\chi^{(3)}$ as high as $(6 \pm 0.58) \times 10^{-19}$ m$^2$/v$^2$. Additionally, we make the argument in favor of PECVD based deposition of SRN over other deposition methods as a technique readily available in CMOS process flows for realizing highly nonlinear SRN films for electro-optic switching applications.

## Theory

As introduced in the previous section, plasma-dispersion based switching is commonly utilized for switching applications in silicon; however, this technique produces, an often un-desirable, change in the imaginary part of the refractive index as well as the real part [8]. An alternative to plasma dispersion is to utilize the Pockels effect [9], such as is exploited by Lithium Niobate modulators. However, due to central crystal symmetry silicon does not possess a $\chi^{(2)}$. As an alternative to this, the DC-Kerr effect has recently been demonstrated in Silicon utilizing a p-i-n junction configuration [7]. This has been demonstrated to be an effective method for realizing electro-optic modulation in silicon. However, its realization has currently been limited to p-i-n junction configurations as a means of overcoming the challenges of engineering efficient overlap of the electric field with the optical mode within a semiconductor. The i-th component of the refractive index modulation, $\Delta n_i$ due to DC induced Kerr effect is given by,

$$\Delta n_i = \frac{3\,\Gamma_{SRN}}{2} \sum_{j,k,l} \frac{\chi^{(3)}_{ijkl}}{n_{l,eq}} E^{dc}_j E^{dc}_k ,\quad (1)$$

where $\Gamma_{SRN}, n_{l,eq}, \chi^{(3)}_{ijkl}, E^{dc}_j$, and $E^{dc}_k$ represent the overlap factor, the unperturbed material index of the "lth" polarization, "ijkl" tensor component of the $\chi^{(3)}$, the jth and kth component of the applied electric field, respectively. Examining equation 1 we observe an expression for the change in refractive index '$\Delta n_i$' caused by the DC Kerr effect, also known as quadratic electro-optic effect, upon application of an external electric field to a material possessing a third order nonlinear susceptibility ($\chi^{(3)}$). The utilization of electrodes in a top-down configuration, with a grounded substrate, results in an applied electric field ($E^{dc}_j$) which is predominately aligned normal to the thin SRN film and along with isotropic material symmetry allows equation 1 to be reduced to equation 2 (see Fig 1(a)). Details on the tensorial properties of $\chi^{(3)}$ for isotropic materials can be found in the supplementary material section S.1., yielding for our configuration an expression:

$$\Delta n_i = \frac{3\,\Gamma_{SRN}}{2} \sum_{i,j} \frac{\chi^{(3)}_{ijji}}{n_{i,eq}} E^{dc^2}_j \qquad (2)$$

It should further be noted that for SRN's material class only certain tensor components will be non-zero. It is evident that with this approach we can achieve electro-optic switching in any material platform regardless of its crystal symmetry, as long as the given material has a large enough combination of $\chi^{(3)}$ and high electric breakdown field strength. However, in SRN the overall change in the material index is often a combination of both second and third order contributions, due to the Pockels and DC-Kerr effects respectively, even if the second order contributions are often small [5]. As such an accurate estimate of the $\chi^{(3)}$ requires consideration for the contributions of both these terms due to the presence of a non-zero $\chi^{(2)}$. For these reasons, SRN becomes a very attractive candidate. Specifically, it has been shown in literature that SRN thin films can exhibit very high third order nonlinear susceptibilities, even larger than that of silicon itself [5], have a higher breakdown field [10], while remaining a low loss dielectric waveguiding material [11,12]. It is for this reason that SRN deserves strong consideration as a candidate platform for electro-optic switching. In the following, we carry out electro-optic measurements of the nonlinear optical response of these films used as a wave guiding material.

## Design and Fabrication

In this work we choose to utilize a bus-coupled racetrack ring resonator and a TM- polarized optical mode to carry out electro-optic characterization of the all-normal components of the SRN film's susceptibility tensors (see Fig 1(a)). Details of the PECVD SRN film deposition are outlined in our previous work [13]. In parallel we also prepared samples of Au-SRN-Au capacitors in order to confirm the safe operating range of electric fields for these films. From these measurements we confirm our films to be safe for application of fields up to 1.2e8 v/m, which is in line with electric field breakdown strength measurements that have been carried out on SRN films with similar silicon compositions [14] and approximate our films RF permittivity to be 9.0578. Ellipsometric measurements performed at a wavelength of 800nm confirm our films to increase in refractive index when they undergo rapid thermal annealing (RTA) at 300°C [15]. For more information see supplementary sections S.2 and S.3 respectively. Our device architecture uses a 320nm thick PECVD SRN device layer on top of a 3 $\mu m$ wet thermal oxide on a silicon handling wafer. The device consists of a point-coupled racetrack ring resonator with a bend radius of 45 $\mu m$ and straight arm sections of 250 $\mu m$. All such resonators have a waveguide width of 450nm and coupling gaps ranging from 100nm to 400nm. The device geometry is written by electron beam lithography in a 400nm thick fox-16 soft mask followed by dry etching using reactive ion etching in an Oxford P100 etcher. After etching, the remaining fox-16 is removed using 1:10 Buffered Oxide Etchant diluted in D.I. water. The devices are then top clad with a 1 $\mu m$ thick layer of PECVD $SiO_2$. Electrode traces 30 $\mu m$ wide with a 10 $\mu m$ separation are then patterned on top of the top clad using an AZ1512/SF9 bi-layer soft mask and DC sputtering of gold in a Denton 635 sputtering system. These samples then undergo RTA at 300°C in a forming gas ambient for 15 mins. The devices are then diced to expose the edge facets of the waveguides.

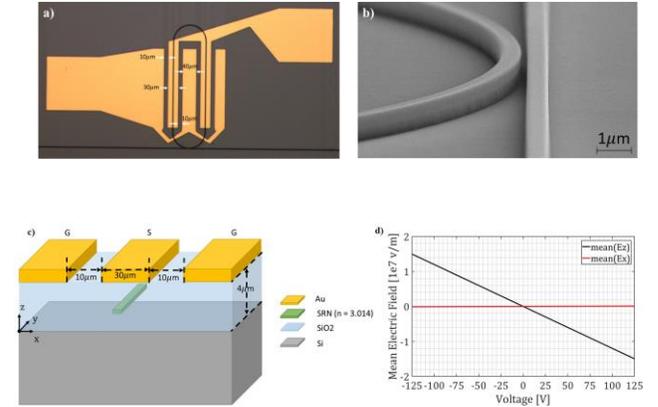

Fig. 1. (a) Optical microscope image of our point coupled ring resonator with electrode width and spacing marked with white arrows and text overlay. (b) Scanning Electron Microscope image of the point coupler region of a racetrack ring resonator with the FOX-16 on top of the post etched waveguides. (c) Schematic breakdown of the phase shifter design shown as a cutout. Note that the substrate of the sample is also grounded. (d) Simulation Results for the mean value of the Applied Field $E_x$ and $E_z$ components versus voltage inside the waveguide core.

The electrode design plays an important role in our study, as we wish to be able to attribute our measurements to specific tensor components, in order to do this, we need to ensure a uniform and directional distribution of the applied DC electric field. Figure 1(c) shows the electrode design while figure 1(d) shows simulation results demonstrating that by using a wide central trace along with grounding the sample substrate the applied field distribution can be made to have as much as a 3 order of magnitude difference between the in-plane and out-of-plane applied fields. This allows us to uniquely attribute our electro-optic measurements of the TM-like polarized optical mode to the $\chi^{(3)}_{3333}$ tensor component, see the supplementary section S.4 for more information.

## Characterization Technique and Experimental Results

For characterization we use a fiber coupled input, free-space output setup with a tunable Agilent 8164B CW source which has a wavelength span of 1465nm to 1575nm [5]. Electrical probes are used to contact the electrical pads (see figure 1(a)) applying voltage from a Keithly Source Meter 2400 with a maximum voltage range of ±210V.

To characterize the DC-Kerr effect, transmission spectra measurements are taken as a function of applied voltage. Using these spectral measurements, the shift in resonant wavelength versus voltage can be extracted. This measurement of resonant wavelength shift ($\Delta\lambda_{res}$) versus voltage can then be related to the change in effective index ($\Delta n_{eff}$) on a first order approximation using equation (3) below as a ratio of the group index ($n_g$) and the resonant wavelength ($\lambda_{res}$) scaled by the modulated length of the ring ($L_{mod}$) as a fraction of total length of the ring ($L_{total}$), 440 $\mu m$ and 782.74 $\mu m$ respectively [16,17]. An important note is that as was discussed previously, our choice to use electrodes in a top-down configuration ensures we have an all-vertical applied field. Thus, when these measurements are carried out for a TM-like optical mode this allows us to attribute our measurements entirely to the all normal type $\chi^{(2)}_{333}$ and $\chi^{(3)}_{3333}$.

$$\frac{\Delta n_{eff}}{n_g} \frac{L_{mod}}{L_{total}} = \frac{\Delta\lambda_{res}}{\lambda_{res}} \qquad (3)$$

The results of these measurements are shown in figure 2 (a) - (c).

These measurements are used to determine $\chi^{(2)}_{333}$ and $\chi^{(3)}_{3333}$ by fitting for the change in effective refractive index as a function of applied field using Lumerical mode simulations to solve for the effective refractive index of the guided mode as function of change in spatial dependent material index for the waveguide core. This spatial dependent material index is derived from modeling of the applied field within the waveguide core as a function of applied field using Lumerical Device. In this way a best fit optimization is done to determine $\chi^{(2)}_{333}$ and $\chi^{(3)}_{3333}$ from the measurements, the results of which can be seen in figure 2(c).

The propagation loss of the circulating mode can be seen extracted from figure 2(a) yields an in-waveguide propagation loss of 3.41dB/cm for the TM-like polarization. This is done by performing a best fit optimization in order to fit the experimental results to the analytical formula for an all-pass ring resonator shown in [16]. The results of the TM-like active optical measurements of resonant wavelength shift versus voltage can be seen in figure 2(b) with figure 2(c) showing the resulting change in effective index versus voltage using equation 3. Figure 2(c) shows the fit for $\chi^{(2)}_{333}$ and $\chi^{(3)}_{3333}$ resulting in values of (14.5 ± 1.4) pm/v and (6 ± 0.58)x10$^{-19}$ m$^2$/v$^2$ respectively. In figures 2(b) and (c) we observe a clear quadratic dependence of the change in resonance wavelength, and change in effective index, on the applied voltage with the minima point displaced from the origin due to the presence of a non-zero $\chi^{(2)}_{333}$ in as grown SRN films. The uncertainty in the $\chi^{(2)}$ and $\chi^{(3)}$ is primarily due the relative wavelength accuracy of the tunable Agilent 8164B CW source used in our measurements. The relative wavelength accuracy of our Agilent 8164B CW source is ± 3pm, resulting in an uncertainty in $\Delta n_{eff}$ of ± 1.35x10$^{-5}$.

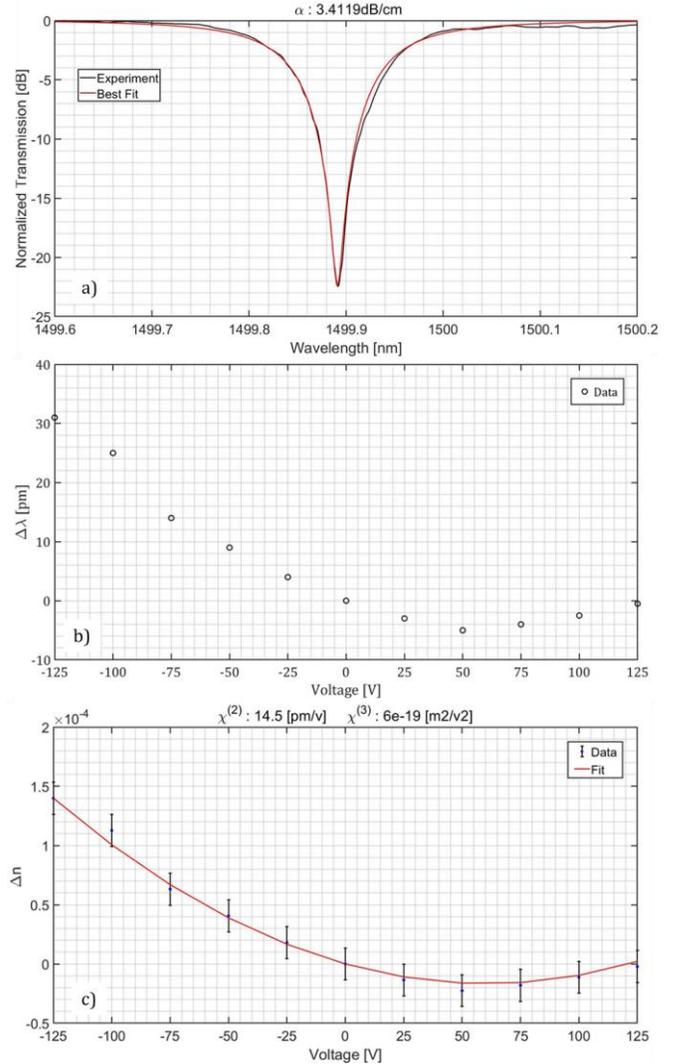

Fig. 2. Above the top cladding there are electrodes 220um in length on each of the two straight arm sections. All measured data is operated in the TM polarization for a 320nm thick 450nm wide SRN waveguide. (a) Measurement of the transmission versus wavelength along with the best fit for a single resonance showing the propagation loss of the circulating mode. (b) Shift in resonance wavelength in pico-meters versus applied voltage. (c) Change in effective index versus applied voltage along with the best fit results resulting in a $\chi^{(2)}_{333}$ of (14.5 ± 1.4) pm/v and a $\chi^{(3)}_{3333}$ of (6 ± 0.58)x10$^{-19}$ m$^2$/V$^2$.

Comparing this results to other works in literature [12,18-22] we find our SRN film to be in line with predictions made in the literature based on the relationship between a SRN film and the $\chi^{(3)}$ it will exhibit [4]. Additionally, we find the $\chi^{(3)}$ of our films to exceed that of Silicon. Furthermore, the clear shift in the minimum point in change in refractive index versus voltage confirms the presence of a combination of second and third order nonlinearities as shown in the fitting results.

## Conclusion

In this work we have presented a PECVD based SRN film with a refractive index of 3.01896 at 1500nm, achieved an in-waveguide propagation loss of 3.41dB/cm for the TM-like polarized optical mode. We have also shown that SRN films, in contrast to highly nonlinear materials such as BaTiO$_3$ [23], can retain a low RF permittivity of 9.06 when deposited using PECVD. Additionally, in this work we have maintained all processing steps at a temperature of 350°c or below. While it is true that lower-loss highly nonlinear SRN films have been demonstrated in literature, existing work has primarily focused on LPCVD and ICP-CVD [12,24,25]. In comparison to these techniques PECVD has a major advantage, readily available compatibility with CMOS process flows and maintains a low thermal budget compared to LPCVD, which are typically deposited at temperatures as high as 800°c and undergo long annealing processes at temperatures as high as 1200°c.

Using the DC-Kerr effect we have experimentally measured the $\chi^{(3)}_{3333}$ of our SRN film to be $(6 \pm 0.58) \times 10^{-19}$ m$^2$/v$^2$ for the TM-like polarized optical mode in the presence of a vertical applied field. The DC-Kerr effect in SRN is used as an optical phase shifter for tuning of a ring-resonator device. This technique can offer an alternative mechanism for employing optical phase shifters in SRN films. Furthermore, PECVD SRN is a highly tunable material which allows a designer to control its refractive index [4], thermo-optic coefficient [24], as well as second and third order nonlinear susceptibilities [3-5] while maintaining low loss and two photon absorption with breakdown field strengths superior to that of silicon [4]. Additionally, when processed using PECVD, a low thermal budget can be maintained using a deposition technique readily available in CMOS process flows. As such, SRN is a very promising material platform which deservers further exploration for on chip applications towards highspeed electro-optic switches, analog transmitters and microwave photonics.

**Funding.** This work was supported by the Defense Advanced Research Projects Agency (DARPA) DSO NLM and NAC programs, the Office of Naval Research (ONR), the National Science Foundation (NSF) grants DMR-1707641, CBET-1704085, NSF ECCS-180789, NSF ECCS-190184, NSF ECCS-2023730, NSF ECCS-190184, the Army Research Office (ARO), the San Diego Nanotechnology Infrastructure (SDNI) supported by the NSF National Nanotechnology Coordinated Infrastructure (grant ECCS-2025752ECCS-1542148), the Quantum Materials for Energy Efficient Neuromorphic Computing-an Energy Frontier Research Center funded by the U.S. Department of Energy (DOE) Office of Science, Basic Energy Sciences under award # DE-SC0019273; Advanced Research Projects Agency Energy (LEED: A Lightwave Energy-Efficient Datacenter), and the Cymer Corporation.

**Acknowledgment**. We thank all of UCSD's nano3 cleanroom staff and Dr Maribel Montero for their assistance with sample fabrication.

**Disclosures**. The authors declare no conflicts of interest.

**Supplemental Document.** See supplementary for supporting content.